\documentstyle[prd,aps]{revtex}
\begin{document}

\draft
\title{Superfluid phases of quark matter. III. Supercurrents and vortices}
\author{Kei Iida$^{1,2,3}$ and Gordon Baym$^{1,2}$}
\address{$^{1}$Department of Physics, University of Illinois at
Urbana-Champaign, 1110 West Green Street, Urbana, Illinois 61801-3080}
\address{$^{2}$Department of Physics, University of Tokyo,
7-3-1 Hongo, Bunkyo, Tokyo 113-0033, Japan}
\address{$^{3}$The Institute of Physical and Chemical Research (RIKEN), 2-1
Hirosawa, Wako, Saitama 351-0198, Japan}
\date{\today}
\maketitle
\begin{abstract}

    We study, within Ginzburg-Landau theory, the responses of three-flavor
superfluid quark-gluon plasmas to external magnetic fields and rotation, in
both the color-flavor locked and isoscalar color-antitriplet diquark phases
near the critical temperature.  Fields are incorporated in the gradient energy
arising from long wavelength distortions of the condensate, via covariant
derivatives to satisfy local gauge symmetries associated with color and
electric charge.  Magnetic vortex formation, in response to external magnetic
fields, is possible only in the isoscalar phase; in the color-flavor locked
phase, external magnetic fields are incompletely screened by the Meissner
effect.  On the other hand, rotation of the superfluid produces vortices in
the color-flavor locked phase; in the isoscalar phase, it produces a London
gluon-photon mixed field.  We estimate the coherence and Meissner lengths and
critical magnetic fields for the two phases.

\end{abstract}

\section{Introduction}

    Dense quark matter at low temperatures is expected to be in a BCS-paired
superfluid state induced by attractive coupling in the color-${\bf{\bar 3}}$
channel, with the pairs having zero total angular momentum, $J$
\cite{barrois,BL,reviews}.  Because of the rich internal degrees of freedom of
quarks, two $J=0$ pairing states with condensates antisymmetric in color and
flavor have emerged as energetically favorable candidates, a two-flavor
color-antitriplet ``2SC" (or particularly for $u$ and $d$ quarks, isoscalar)
state, and for massless $u$, $d$, and $s$ quarks, a color-flavor locked state
\cite{ARW,EHHS,tom,I}.  The latter is the most stable for three flavors of
massless quarks in the weak coupling limit, both at zero temperature, $T$, and
near the critical temperature $T_c$ [I].

    In this paper we study the response of BCS-paired states of quark matter
to external magnetic fields and rotation, as would be experienced by
superfluid quark matter if present in neutron stars.  We find that the
isoscalar paired state, if a Type II superconductor, behaves analogously to
that of protons in a neutron star, forming vortices in response to the
magnetic field, and forming a weak London magnetic field dominated by the
gluonic component, in response to rotations.  On the other hand, the
color-flavor locked phase forms vortices in response to rotation, analogously
to superfluid neutrons in a neutron star or to rotating superfluid $^3$He, but
simply tries to expel, via the Meissner effect, a fraction of the external
magnetic field \cite{fluxdiff}; gradients of the order parameter do not induce
supercurrents, and magnetized vortices are topologically unstable, in contrast
to the situation in the isoscalar phase.  Table \ref{summary} summarizes our
main conclusions on the responses and corresponding energies.  In neutron
stars, the presence of the isoscalar phase would affect the magnetic fields
of the star via the combined effects of Meissner screening, London
magnetic fields, and possible magnetic vortices \cite{BS,ABR2,SBSV}, while the
color-flavor locked phase could play a role \cite{ABR3} in the pinning and
depinning of rotational vortices that give rise to pulsar glitches \cite{BEL}.

    To investigate the nature of vortices and supercurrents induced by
external magnetic fields and rotation in pairing states near the critical
temperature, we extend the general Ginzburg-Landau theory constructed in I by
adding to the homogeneous free energy the energy arising from long wavelength
gradients of the order parameters.  The gradient energy, which is applicable
over spatial scales larger than the zero-temperature coherence length $\sim
T_c^{-1}$, is proportional to the superfluid baryon density $n_s$, which we
obtain explicitly in weak coupling by calculating the normal baryon density
$n_n$.  We then incorporate external magnetic fields and rotation via
covariant derivatives and transformations to the rotating frame, respectively.
The Ginzburg-Landau free energy including color and electromagnetic gauge
fields was analysed by Bailin and Love \cite{BL} and by Blaschke and Sedrakian
\cite{BS} for the isoscalar channel in the weak coupling limit.  The present
study encompasses both color-flavor locked and the isoscalar condensation of
arbitrarily coupled systems under external magnetic fields and rotation.

    We derive the effects of uniform external magnetic fields on the
condensates from the static Maxwell's equations in the medium, constructed by
extremization of the Ginzburg-Landau free energy with respect to the gluonic
and photonic gauge fields; the supercurrent sources include terms induced by
the gradients of the order parameters, as in ordinary superconductors
\cite{dG1}.  The response of the system to uniform magnetic fields is
characterized by Meissner screening and free propagation of gluon-photon mixed
fields, together with possible formation of magnetized vortices
\cite{BS,ABR2,SBSV}.  In clarifying the magnetic structures of the
superconductor, we calculate the Ginzburg-Landau coherence length and the
penetration depth, as well as the various critical fields.  We treat all the
gauge fields as averaged quantities and ignore fluctuations around their mean
values, which cause the normal-superfluid transition to become first order, as
clarified in Refs.\ \cite{HLM,BG,BL}.

    Magnetic fields and rotation have negligible effect on the equilibrium
phase diagram near $T_c$.  Neither of these disturbances affects the
degeneracy of the states belonging to a given phase because they do not couple
with the orientations of the order parameters in color and flavor space.  This
is a contrast to the case of superfluid $^3$He in which external magnetic
fields and currents act to fix the orientations of the order parameters in
spin and orbital space, respectively, and hence remove such degeneracy
\cite{helium}.  In addition, the color neutrality conditions, discussed in I,
remain unchanged, since neither magnetic fields or rotation alter color charge
densities via gluon self-couplings; we do not take these conditions into
account in this paper.  While both rotation and magnetic fields yield small
positive energy corrections that are more favorable to the isoscalar phase
than the color-flavor locked phase, such corrections are considerably smaller
than the condensation energies except in a negligibly small temperature range
near $T_c$.

    In Sec.\ \ref{sec:grad}, we construct the gradient energy and the
supercurrents for color and flavor antisymmetric pairing with $J=0$, and
determine the superfluid baryon density and superfluid momentum density in
terms of the order parameter.  In Sec.\ \ref{sec:CFL} we examine the
responses of a color-flavor locked condensate to external magnetic fields and
rotation, and in Sec.\ \ref{sec:IS}, the corresponding responses of isoscalar
condensates.  Our conclusions are presented in Sec.\ \ref{sec:conclusion}.
The appendix briefly summarizes crucial results from I. \, We work in units
$\hbar=c=k_B=1$.

\begin{table}
\caption{Responses of condensates to magnetic fields and rotation
  [corresponding energies]}
\label{summary}
\begin{tabular}{lll}
 &  Magnetic fields  & Rotation \\
\hline
Color-flavor locking &  Partial screening, no vortices
& $U(1)$ vortices \\
 & [Eq.\ (\ref{magcfl})]
 & [Eq.\ (\ref{domegarot})] \\
\hline
Isoscalar &  Partial screening, possible vortices
 & London magnetic field  \\
 & [Eq.\ (\ref{magis})]&
   [Eq.\ (\ref{domegarotis})]
\end{tabular}
\end{table}

\section{Gradient energy}
\label{sec:grad}

    To describe the effects of external magnetic fields and rotation in a
quark superfluid, we first study the gradient term in the Ginzburg-Landau free
energy near $T_c$.  We consider a plasma composed of quarks of three massless
flavors ($uds$) and three colors ($RGB$) at temperature $T$ and baryon
chemical potential $\mu$, as in I, and work in terms of the free energy
density $\Omega = E - TS -\Sigma_{ia} \mu_{ia} n_{ia}$, where $S$ is the
entropy density, and the $\mu_{ia}$ and $n_{ia}$ are the chemical potentials
and densities of quarks of flavor $i$ and color $a$.  In the absence of real
electromagnetic and gluonic fields the free density difference $\Delta
\Omega^{(0)}\equiv\Omega_s^{(0)}-\Omega_n^{(0)}$ between the homogeneous
superfluid and normal phases near $T_c$ is, up to fourth order in the pairing
gap (see appendix),
\begin{eqnarray}
  \Delta\Omega ^{(0)}
  &=& \alpha^{+}{\rm Tr}(\phi_{+}^{\dagger}\phi_{+})_{F}
  +\beta^{+}_{1}[{\rm Tr}(\phi_{+}^{\dagger}\phi_{+})_{F}]^{2}
  +\beta^{+}_{2}{\rm Tr}[(\phi_{+}^{\dagger}\phi_{+})^{2}]_{F}\ .
 \label{domega}
\end{eqnarray}
Here $\phi_{+}$ is the gap matrix for even parity, same chirality, $J=0$
quark-quark pairing (see Eq.~(\ref{delta1})).  The subscript $F$ denotes the
on-shell pairing gap evaluated at $|{\bf k}|=k_F$, the Fermi momentum of the
normal state; hereafter we do not write the $F$ explicitly, but it should be
understood.  We ignore the terms $\Omega_{CN}$ associated with the constraint
of overall color neutrality, discussed in I, since they play no role in the
ensuing discussion.

    In inhomogeneous states $\phi_{+}$ depends not only on the relative pair
momentum, ${\bf k}$, but also on the center-of-mass coordinate, ${\bf r}$, of
the pair.  For wavelengths large compared with the coherence length, $\sim
T_c^{-1}$, the gradient energy assumes the general form to second order in
spatial derivatives of the pairing gap,
\begin{equation}
  \Omega_{g}= \frac12 K_T {\rm Tr}(\partial_l \phi_{+}
   \partial_l \phi_+^{\dagger})\ ,
   \label{omegag}
\end{equation}
where $\partial_l\equiv \partial/\partial r_l$ ($l=1,2,3$).  This structure
follows from the invariance of the grand-canonical Hamiltonian under
special unitary color and flavor rotation and $U(1)$ gauge transformations of
the field operators, $\psi \to e^{i\varphi}U_cU_f\psi$, and thus
$(\phi_{\pm})_{abij}\to e^{-2i\varphi}(\phi_{\pm})_{cdlm}
(U_c^\dagger)_{ca}(U_c^\dagger)_{db}(U_f^\dagger)_{li}(U_f^\dagger)_{mj}$.

    The superfluid mass density is related to the stiffness parameter $K_T$ by
\begin{equation}
 \rho_s = \frac49\mu^2 K_T
 {\rm Tr} (\phi_{+} \phi_{+} ^{\dagger}) \ .
 \label{josephson}
\end{equation}
This relation is basically that obtained by Josephson \cite{josephson},
$\rho_s=A_{\bot}(m/\hbar)^2|\Psi|^2$, for superfluid He-II.  To derive it we
consider the situation, as in Ref.\ \cite{II}, in which the superfluid moves
uniformly with small constant velocity ${\bf v}_s$, with the
normal fluid remaining at rest.  The momentum per baryon in the flow is $\mu
{\bf v}_s$.  In such a situation the phase factor of the gap in the lab frame
transforms by $\phi_+ \to e^{-i{\bf P\cdot r}}\phi_+$, where ${\bf P}$ is the
total pair momentum relative to its value in the lab frame.  Since each pair
carries a baryon number 2/3,
\begin{equation}
   {\bf P} = \frac23 \mu {\bf v}_s.
 \label{pvs}
\end{equation}
Under the transformation $\phi_+ \to e^{-i{\bf P\cdot r}}\phi_+$ the free
energy density transforms by
\begin{equation}
  \Omega\to \Omega - \frac i2 K_T {\bf P}\cdot
  {\rm Tr}(\phi_{+}  \nabla \phi_+^{\dagger}-
  \phi_+^{\dagger} \nabla \phi_{+})\ +
    \frac12 K_T {\bf P}^2  {\rm Tr}(\phi_{+} \phi_+^{\dagger})\ .
\label{galtrans1}
\end{equation}
Comparing this result with the expected transformation of the free energy
density,
\begin{equation}
  \Omega\to \Omega  + {\bf g}_s\cdot {\bf v}_s + \frac12 \rho_s {\bf v}_s^2\ ,
\label{galtrans2}
\end{equation}
and using Eq.~(\ref{pvs}), we derive Eq.~(\ref{josephson}), and in addition
relate the momentum density of the condensate, ${\bf g}_s$, to the order
parameter by
\begin{equation}
   {\bf g}_s = - \frac i3 K_T \mu
  {\rm Tr}(\phi_{+}  \nabla \phi_+^{\dagger}-
  \phi_+^{\dagger} \nabla \phi_{+} )\ =\mu {\bf j}_s\ ,
 \label{gs}
\end{equation}
where ${\bf j}_s$ is the superfluid baryon current.  [The minus sign
originates from the conventional definition of the gap here in terms of the
adjoint spinors rather than the spinors, as in condensed matter physics.]

    In the limit of weak coupling near $T_c$, the stiffness parameter is given
by
\begin{equation}
  K_T = \frac{7\zeta(3)n_b}{16\pi^2 T_c^2 \mu}\ ,
 \label{K}
\end{equation}
where $n_b$ is the baryon density, and $\zeta(3)= 1.202\ldots$.  This
result has been derived by Giannakis and Ren \cite{IH} using finite
temperature diagrammatic perturbation theory in the normal phase for single
gluon exchange interaction, and by Bailin and Love \cite{BL} for a short-range
pairing interaction.  Here we present a simple physical derivation of $K_T$
from Eq.~(\ref{josephson}), by evaluating the superfluid mass density $\rho_s$
near $T_c$ in weak coupling.  We consider the system, as in Leggett's argument
for superfluid $^3$He \cite{helium}, to be contained in an infinitely long
tube which moves uniformly with velocity ${\bf v}$ $(|{\bf v}|\ll c)$ along
the symmetry axis.  We assume that the normal component is in equilibrium with
the moving wall and that the superfluid component stays at rest.  Then the
baryon current density ${\bf j}_n$ of the normal component is related to the
normal baryon density, $n_n=n_b-n_s$, by
\begin{equation}
  {\bf j}_n = n_n {\bf v}\ ,
 \label{jn1}
\end{equation}
and is given in weak coupling in terms of the quasiparticle distribution
function, $f_{\bf k}$, by
\begin{equation}
  {\bf j}_n = 2 \int\frac{d^3 k}{(2\pi)^3} \frac {{\bf k}}{\mu}
  {\rm Tr}f_{\bf k}\ .
 \label{jn4}
\end{equation}
In the present situation, $f_{\bf k}$ is the equilibrium distribution
evaluated in the frame of the moving walls:
\begin{equation}
 f_{\bf k} = f \mbox{\boldmath $($}\varepsilon({\bf k})-{\bf k\cdot v}
    \mbox{\boldmath $)$} =
    \frac{1}{e^{[\varepsilon({\bf k})-{\bf k\cdot v}]/T}+1}\ ;
 \label{distribution}
\end{equation}
here
\begin{equation}
  \varepsilon({\bf k}) = [(|{\bf k}|-\mu/3)^2
                         +\phi_+^\dagger({\bf k})\phi_+({\bf k})]^{1/2}\ ,
\end{equation}
with $\phi_+({\bf k})\equiv\phi_+\mbox{\boldmath $($}\varepsilon({\bf k}),
{\bf k}\mbox{\boldmath $)$}$, is the energy matrix for quark quasiparticles 
[I].  Expanding to linear order in ${\bf v}$ we find
\begin{equation}
  n_n = \frac23
           \int\frac{d^3 k}{(2\pi)^3}
            \frac {{\bf k}^2}{\mu T}
              {\rm Tr}[
             f\mbox{\boldmath $($}\varepsilon({\bf k})\mbox{\boldmath $)$}
          (1-f\mbox{\boldmath $($}\varepsilon({\bf k})\mbox{\boldmath $)$})]\ .
\end{equation}
The integration is dominated by momenta ${\bf k}$ close to the Fermi
surface, where the gap can be taken to be constant.  Expanding near $T_c$ to
second order in the gap we thus obtain
\begin{equation}
  n_n = \left[1-\frac{7\zeta(3)}{36\pi^2 T_c^2}
             {\rm Tr}(\phi_+^\dagger \phi_+) \right]n_b\ ,
 \label{jn3}
\end{equation}
so that
\begin{equation}
  n_s = \rho_s/\mu = \frac{7\zeta(3)}{36\pi^2 T_c^2}
             {\rm Tr}(\phi_+^\dagger \phi_+) n_b\ ,
 \label{nb}
\end{equation}
from which Eq.\ (\ref{K}) follows.

    Real color and electromagnetic gauge fields enter the gradient energy
(\ref{omegag}) via covariant derivatives.  Let us first consider color $SU(3)$
local gauge transformations of the quark spinors via
\begin{equation}
   U_c({\bf r})=\exp[i\lambda^\alpha \varphi_\alpha({\bf r})/2]\ ,
\end{equation}
where the $\varphi_\alpha$ are the local phase angles and the
$\lambda^\alpha$ the Gell-Mann matrices.  The gap matrix $\phi_+(k;{\bf r})$
transforms as $(\phi_+)_{abij} \to
(U_c^\dagger)_{ca}(U_c^\dagger)_{db}(\phi_+)_{cdij}$, and the corresponding
covariant derivative is
\begin{equation}
   D_l\phi_+ \equiv \partial_l \phi_+ + \frac 12 ig
           [(\lambda^\alpha)^* \phi_+ + \phi_+ \lambda^\alpha] A_l^\alpha\ ,
   \label{cd1}
\end{equation}
where $g$ is the qcd coupling constant, and the ${\bf A}^\alpha$ are the color
gauge fields, which transform as $A^\alpha_l \to A^\alpha_l + g^{-1}\partial_l
\varphi_\alpha+f_{\alpha\beta\gamma}A^\beta_l \varphi_\gamma$.
With extension of the covariant derivative (\ref{cd1}) to include
the usual $U(1)$ electromagnetic gauge transformations, which
multiply the quark spinors $\psi_{ai}$ by
\begin{equation}
   U_i({\bf r})=\exp[iq_i\varphi_e({\bf r})]\ ,
\end{equation}
with local phase angle $\varphi_e$, the full covariant derivative is then
\cite{gorbar}
\begin{equation}
   D_l\phi_+ \equiv \partial_l \phi_+ + \frac 12 ig
           [(\lambda^\alpha)^* \phi_+ + \phi_+ \lambda^\alpha] A_l^\alpha
           +ie Q \phi_+ A_l\ ,
   \label{cd2}
\end{equation}
where ${\bf A}$ is the electromagnetic gauge field, $e$ is unit of the
electric charge, and $Q_{abij}=\delta_{ab}(q_i+q_j)$ is the electric charge
matrix of the pair.  The full gradient energy is then
\begin{equation}
  \Omega_{g}= \frac12 K_T {\rm Tr}[D_l \phi_{+}
   (D_l \phi_+)^{\dagger}]\ ,
   \label{omegag1}
\end{equation}
with $D_l$ given by Eq.\ (\ref{cd2}).  This form is invariant under
transformations $(\phi_+)_{abij} \to  U_i^* U_j^* (U_c^\dagger)_{ca}
(U_c^\dagger)_{db}(\phi_+)_{cdij}$.

    We now write the free energy densities, $\Omega_n$ and $\Omega_s$, of the
normal and superfluid phases near $T_c$ in the presence of an external
magnetic field, ${\bf H}_{\rm ext}$, neglecting the response of the normal
component to ${\bf H}_{\rm ext}$.  First,
\begin{equation}
  \Omega_n=F_n+\frac12 |{\bf H}_{\rm ext}|^2\ ,
  \label{domega11}
\end{equation}
where $F_n$ is the free energy of the normal state at ${\bf H}_{\rm
ext}=0$.  To construct the free energy density of the superfluid state, we add
to the homogeneous free energy (\ref{domega}), the gradient energy density
(\ref{omegag1}), and the energy densities of the color and electromagnetic
fields induced by ${\bf H}_{\rm ext}$, to obtain
\begin{eqnarray}
  \Omega_s &=&
   F_n+
   \alpha^{+}{\rm Tr}(\phi_{+}^{\dagger}\phi_{+})
  +\beta^{+}_{1}[{\rm Tr}(\phi_{+}^{\dagger}\phi_{+})]^{2}
  +\beta^{+}_{2}{\rm Tr}[(\phi_{+}^{\dagger}\phi_{+})^{2}]
  \nonumber \\
  & & +\frac12 K_T {\rm Tr}[D_l \phi_{+}
   (D_l \phi_+)^{\dagger}]\ + \frac 14 G_{lm}^\alpha G_{lm}^\alpha
   +\frac 14 F_{lm}F_{lm}\ ,
  \label{domega1}
\end{eqnarray}
where
\begin{equation}
   G_{lm}^\alpha = -\partial_l A_m^\alpha + \partial_m A_l^\alpha
      -g f_{\alpha\beta\gamma}A_l^\beta A_m^\gamma
  \label{glm}
\end{equation}
and
\begin{equation}
   F_{lm} = -\partial_l A_m + \partial_m A_l
  \label{flm}
\end{equation}
are the field tensors.

    Extremization of $\int d^3 {r} \Delta\Omega$, where $\Delta\Omega\equiv
\Omega_s-\Omega_n$, with respect to $\phi_+^\dagger$ yields the gap
equation
\begin{equation}
 -\frac 12 K_T D_l (D_l \phi_+) \alpha^+ \phi_+
   +2\beta_1^+ [{\rm Tr}(\phi_+^\dagger \phi_+)] \phi_+
   +2\beta_2^+ \phi_+ \phi_+^\dagger \phi_+ =0\ .
 \label{gapeq}
\end{equation}
The field equations for the macroscopic fields ${\bf A}^\alpha$ and ${\bf
A}$, found by extremizing $\int d^3 {r} \Delta\Omega$ with respect to the
fields, are:
\begin{eqnarray}
  \lefteqn{
  \partial_m G_{ml}^\alpha + g f_{\alpha\beta\gamma }A_m^\beta G_{ml}^\gamma
  } \nonumber \\
 &=& - \frac12 K_T g {\rm Im}\left\{{\rm
     Tr}\left[\left((\lambda^\alpha)^*
                          \phi_+ + \phi_+ \lambda^\alpha\right)^\dagger
                          \partial_l \phi_+\right] \right\}
 \nonumber \\
 & & - \frac14 K_T g^2 A_l^\beta {\rm Re}\left\{{\rm Tr}\left[
            \left((\lambda^\alpha)^* \phi_+ + \phi_+ \lambda^\alpha\right)
            \left((\lambda^\beta)^* \phi_+
                   + \phi_+ \lambda^\beta\right)^\dagger
                 \right] \right\}
 \nonumber \\
& & - \frac12 K_T ge A_l {\rm Re}\left\{{\rm Tr}\left[
   \left(   (\lambda^\alpha)^* \phi_+ + \phi_+ \lambda^\alpha \right)
      Q\phi_+^\dagger  \right] \right\}
       \equiv J_l^\alpha
 \label{cmax}
\end{eqnarray}
and
\begin{eqnarray}
  \partial_m F_{ml} &=&
   - K_T e {\rm Im}[{\rm Tr}(Q\phi_+^\dagger \partial_l \phi_+)]
   - K_T e^2 A_l {\rm Tr}(Q\phi_+ Q\phi_+^\dagger)
 \nonumber \\
 & & - \frac12 K_T ge A_l^\alpha {\rm Re}
  \left\{{\rm Tr}\left[Q\phi_+ \left((\lambda^\alpha)^* \phi_+
                   + \phi_+ \lambda^\alpha\right)^\dagger
                   \right] \right\}  \equiv J_l\ ,
 \label{emmax}
\end{eqnarray}
where ${\bf J}^\alpha$ and ${\bf J}$ are the color and electric
supercurrent densities, respectively.  The Maxwell equations (\ref{cmax}) and
(\ref{emmax}) with uniform external fields determine the structures of the
supercurrents and vortices, as we discuss in Secs.\ \ref{sec:CFL} and
\ref{sec:IS}.

    For later analysis it is useful to write down the solution of these 
equations in the presence of an externally applied electromagnetic transverse 
field, ${\bf A}_{\rm ext}({\bf r})=e^{i{\bf q \cdot r}} {\bf A}_{\rm ext}
({\bf q})$, where ${\bf q\cdot A}_{\rm ext}({\bf q}) = 0$.  The supercurrents,
${\bf J}$ and ${\bf J}^\alpha$, induced by this potential in turn induce 
fields, ${\bf A}_{\rm ind}$ and ${\bf A}^\alpha_{\rm ind}$.  We assume that 
the external potential ${\bf A}_{\rm ext}$ is sufficiently weak and nearly 
uniform (${\bf q}\to 0$) that the order parameter remains essentially 
homogeneous and the self-couplings of the induced gluon fields are negligible.
Up to linear order in ${\bf A}_{\rm ext}$, the induced fields may be 
calculated from Eqs.\ (\ref{cmax}) and (\ref{emmax}), with $G_{lm}^\alpha$ and 
$F_{lm}$ including the induced fields.  The supercurrents ${\bf J}$ and 
${\bf J}^\alpha$ depend on the total fields,
\begin{equation}
   {\bf A}={\bf A}_{\rm ext}+{\bf A}_{\rm ind}\ ,\quad
   {\bf A}^\alpha={\bf A}^\alpha_{\rm ind}\ ,
\end{equation}
which are given by
\begin{eqnarray}
  {\bf A}({\bf q})&=&\left\{1+\frac{K_T e^2}{q^2}{\rm Tr}(Q\phi_+
                    Q\phi_+^\dagger)
      -[(\varepsilon^T)^{-1}]_{\alpha\beta} \left(\frac{K_T ge}{q^2}\right)^2
  \right.
   \nonumber \\  & &
  \times \frac14 {\rm Re}
         \left[{\rm Tr}\left(Q\phi_+ \left((\lambda^\alpha)^* \phi_+
                   + \phi_+ \lambda^\alpha\right)^\dagger
                   \right) \right]
   \nonumber \\
    & & \left. \times {\rm Re}
         \left[{\rm Tr}\left( \left(
         (\lambda^\beta)^* \phi_+
                   + \phi_+ \lambda^\beta \right)
                    Q\phi_+^\dagger
                   \right) \right]\right\}^{-1}
       {\bf A}_{\rm ext}({\bf q})
   \label{res1}
\end{eqnarray}
and
\begin{equation}
   {\bf A}^\alpha({\bf q})=
    -\frac12 [(\varepsilon^T)^{-1}]_{\alpha\beta}\frac{K_T ge}{q^2}
        {\rm Re}\left[{\rm Tr}\left( \left(
        (\lambda^\beta)^* \phi_+
         + \phi_+ \lambda^\beta  \right)
         Q\phi_+^\dagger \right) \right]
      {\bf A}({\bf q})\ ,
   \label{res2}
\end{equation}
where the transverse color dielectric function,
\begin{eqnarray}
 \varepsilon_{\alpha\beta}^T&=&\delta_{\alpha\beta}
   \nonumber \\    & &
       +\frac{K_T g^2}{4q^2} {\rm Re}\left\{{\rm Tr}\left[
            \left((\lambda^\alpha)^* \phi_+ + \phi_+ \lambda^\alpha\right)
            \left((\lambda^\beta)^* \phi_+
                   + \phi_+ \lambda^\beta\right)^\dagger
                 \right] \right\} \ ,
\end{eqnarray}
describes the response of the current of color $\alpha$ to the field of
color $\beta$.  In deriving this equation from Eq.~(34) of II, we have
written the long wavelength limit of the transverse screened correlation
function ${\tilde\chi}_T^{\alpha\beta}({\bf k},0)$ in terms of the gap as
\begin{equation}
  {\tilde\chi}_T^{\alpha\beta}({\bf k}\to 0,0)
  = \frac14 K_T g^2 {\rm Re}\left\{{\rm Tr}\left[
            \left((\lambda^\alpha)^* \phi_+ + \phi_+ \lambda^\alpha\right)
            \left((\lambda^\beta)^* \phi_+
                   + \phi_+ \lambda^\beta\right)^\dagger
                 \right] \right\}\ .
\end{equation}

    In the succeeding sections we analyze the structures of inhomogeneous
color-flavor locked and isoscalar condensates.  These condensates are
antisymmetric in color and flavor, with a gap
\begin{equation}
  \phi_+ = \epsilon_{ijh}\epsilon_{abc} ({\bf d}_c)_h\ ,
  \label{phi1}
\end{equation}
where $({\bf d}_c)_h$ is a complex field representing the pairing gap
between two quarks in state $[(|ab\rangle-|ba\rangle)/\sqrt2] \times
[(|ij\rangle-|ji\rangle)/\sqrt2]$ with $a \neq b \neq c$ and $i \neq j \neq
h$.  With (\ref{phi1}), Eqs.\ (\ref{domega1}) and (\ref{domega11}) for
$\Omega_s$ and $\Omega_n$ lead to a thermodynamic potential density
difference near $T_c$
\begin{eqnarray}
   \Delta\Omega &=&   {\bar\alpha}\lambda
         +(\beta_{1}+\beta_{2}\Upsilon)\lambda^{2}
   \nonumber \\ & &
        +K_T {\rm Tr}[D_l \Phi
           (D_l \Phi)^{\dagger}] + \frac 14 G_{lm}^\alpha G_{lm}^\alpha
        +\frac 14 F_{lm}F_{lm}-\frac12 |{\bf H}_{\rm ext}|^2\ ,
   \label{domega2}
\end{eqnarray}
where $\Phi_{abi} \equiv \epsilon_{abc}({\bf d}_c)_i$.  The homogeneous
part is that in Eq.\ (\ref{domega3}).

    We turn now to the properties of the color-flavor locked and isoscalar
condensates, with respective gap matrices (\ref{phicfl}) and (\ref{icat1}),
in the presence of external magnetic fields and rotation.

\section{Color-flavor locking}
\label{sec:CFL}

    In the color-flavor locked phase, vortices associated with the response of
the system to magnetic fields are topologically unstable.  Rather, the system
acts as an imperfect diamagnet.  On the other hand, rotation of the
color-flavor locked phase produces vortices associated with the baryon $U(1)$
symmetry breaking, analogous to the vortices in rotating chargeless
superfluids such as liquid He II \cite{rotheII}, superfluid $^3$He
\cite{rothe3}, Bose-Einstein condensates of alkali atoms \cite{rotbec}, and
neutron superfluids \cite{bpp,sauls}.

    We consider the color-flavor locked phase described by the gap
(\ref{phicfl})
\begin{equation}
  (\phi_{+})_{abij}=\kappa_{A}
  (\delta_{ai}\delta_{bj}-\delta_{aj}\delta_{bi})\ .
 \label{phicfl1}
\end{equation}
In the absence of magnetic and color magnetic fields the gap equation
becomes
\begin{equation}
  {\bar\alpha}\kappa_A
          + 6{\bar\beta}_{\rm CFL}|\kappa_A|^2 \kappa_A
           -2K_T \nabla^2 \kappa_A =0\ ,
\label{gapeqnkappa}
\end{equation}
where $\bar\beta_{\rm CFL}=\beta_1+\beta_2/3$.  In the London limit, in
which the scale $L$ of the spatial variation is much larger than the
Ginzburg-Landau coherence length,
\begin{equation}
   \xi_{\rm CFL} = \left(2K_T/|{\bar\alpha}|\right)^{1/2}\ ,
  \label{clcfl}
\end{equation}
the relative variation of the magnitude of $\kappa_A$ is of order $\xi_{\rm 
CFL}^2/L^2$ smaller than the relative variation of the phase of $\kappa_A$ and
can be neglected.  To see this we write $\kappa_A =e^{i\varphi_A}|\kappa_A|$.
In the homogeneous limit, $\varphi_A$ can be taken to be zero, and the
magnitude of the gap is given by
\begin{equation}
  {\bar\alpha}
          + 6{\bar\beta}_{\rm CFL}|\kappa_A|^2 = 0\ .
  \label{alphabeta}
\end{equation}
Thus in an inhomogeneous situation the variation of the magnitude of the
gap is related to the variation of its phase by
\begin{equation}
  \frac{\delta|\kappa_A|}{|\kappa_A|}
           = \frac12 \xi_{\rm CFL}^2 e^{-i\varphi_A}
             \nabla^2 e^{i\varphi_A}\ .
  \label{deltakappa}
\end{equation}
Equation~(\ref{alphabeta}) implies the homogeneous phase condensation energy
\begin{equation}
    -\Delta \Omega
    =  \frac{{\bar\alpha}^2}{4{\bar\beta}_{\rm CFL}}\ .
 \label{econd}
\end{equation}

\subsection{Response to magnetic fields}
\label{subsec:CFLMAG}

    To determine the response of the condensate to an external electromagnetic
vector potential we substitute the gap (\ref{phicfl1}) into the field
equations (\ref{cmax}) and (\ref{emmax}), and neglecting gradients of the
magnitude of $\kappa_A$, we find the color and electromagnetic supercurrent
densities:
\begin{equation}
   {\bf J}^\alpha= - 2K_T g^2 |\kappa_A|^2 {\bf A}^{\alpha}\ ,~~~
              \alpha\neq3,8\ ,
   \label{j1cfl}
\end{equation}
\begin{equation}
   {\bf J}^3= - 2 K_T g |\kappa_A|^2 (g{\bf A}^3 +e {\bf A})\ ,
   \label{j3cfl}
\end{equation}
\begin{equation}
   {\bf J}^8= - 2 K_T g |\kappa_A|^2 \left(g{\bf A}^8
               +\frac{e}{\sqrt3} {\bf A}\right)\ ,
   \label{j8cfl}
\end{equation}
\begin{equation}
   {\bf J} =   -2 K_T e |\kappa_A|^2 \left[\frac43 e{\bf A}
           + g\left({\bf A}^3
            + \frac{1}{\sqrt3} {\bf A}^8\right)\right]\ .
   \label{jcfl}
\end{equation}
Note that the current terms induced by the gradients of the phase of
$\kappa_A$ vanish due to the isotropy of the condensate in color and flavor
space, a feature implicit in Refs.~\cite{ABR2,FZ}.  This absence of currents
induced by the gradients of the phase of $\kappa_A$ underlies the response of
the color-flavor locked condensate to magnetic fields and rotation, as we
consider in the present and following subsections.  External
disturbances such as magnetic fields and rotation do not change such isotropy
and hence do not break the degeneracy in energy of the states with order
parameters of the form (\ref{opcfl}).

    The currents associated with color charge $\alpha=3,8$ and electric charge
respond to the electromagnetic gauge field; in turn the electromagnetic
current responds to the gauge fields ${\bf A}^3$ and ${\bf A}^8$.  As in Ref.\
\cite{gorbar}, it is convenient to take linear combinations of the
supercurrents and the gauge fields to diagonalize the response.  The
combinations
\begin{equation}
 \mbox{\boldmath ${\cal J}$}
   \equiv \frac{\sqrt3 g {\bf J} - \sqrt3 e {\bf J}^3 - e {\bf J}^8}
                          {3\sqrt{2}g_3}\ ,~~~
 \mbox{\boldmath ${\cal A}$}
   \equiv \frac{\sqrt3 g {\bf A} - \sqrt3 e {\bf A}^3 - e {\bf A}^8}
                          {3\sqrt{2}g_3}\ ,~~~
   \label{mixcfl}
\end{equation}
\begin{equation}
 \mbox{\boldmath ${\cal J}$}^3
   \equiv \frac{3 g {\bf J}^3 + 4 e {\bf J} + \sqrt3 g{\bf J}^8}
                          {6\sqrt{2}g_3}\ ,~~~
 \mbox{\boldmath ${\cal A}$}^3
   \equiv \frac{3 g {\bf A}^3 + 4 e {\bf A} + \sqrt3 g {\bf A}^8}
                          {6\sqrt{2}g_3}\ ,~~~
   \label{mix3cfl}
\end{equation}
\begin{equation}
 \mbox{\boldmath ${\cal J}$}^8
   \equiv \frac{\sqrt3 {\bf J}^8 - {\bf J}^3}{2}\ ,~~~
 \mbox{\boldmath ${\cal A}$}^8
  \equiv \frac{\sqrt3 {\bf A}^8 - {\bf A}^3}{2}
   \label{mix8cfl}
\end{equation}
satisfy the diagonalized equations
\begin{equation}
 \mbox{\boldmath ${\cal J}$}  = 0\ ,
\end{equation}
\begin{equation}
 \mbox{\boldmath ${\cal J}$}^3
   = -12 K_T g_3^2 |\kappa_A|^2 \mbox{\boldmath ${\cal A}$}^3\ ,
   \label{mixj3cfl}
\end{equation}
and
\begin{equation}
 \mbox{\boldmath ${\cal J}$}^8
   = -2 K_T g^2 |\kappa_A|^2 \mbox{\boldmath ${\cal A}$}^8\ ,
   \label{mixj8cfl}
\end{equation}
where
\begin{equation}
   g_3 = \frac{1}{3\sqrt2} \sqrt{3g^2+4e^2}
\end{equation}
is the coupling constant associated with the field $\mbox{\boldmath ${\cal
A}$}^3$.  Since Eqs.~(\ref{j1cfl}), (\ref{mixj3cfl}), and (\ref{mixj8cfl})
imply that the corresponding fields have a magnetic mass, we see the property
originally shown by Alford, Rajagopal, and Wilczek \cite{ARW}, that eight of
the nine gauge fields are Meissner screened, while the field $\mbox{\boldmath
${\cal A}$}$ remains free.

    The response of a homogeneous color-flavor locked
condensate near $T_c$ to an electromagnetic field is twofold.  Substituting
the supercurrents (\ref{j1cfl})--(\ref{jcfl}) into the linear response
(\ref{res1}) and (\ref{res2}) we obtain
\begin{equation}
  {\bf A}=\frac{q^2+2K_T g^2|\kappa_A|^2}
                    {q^2+12K_T g_3^2 |\kappa_A|^2}
               {\bf A}_{\rm ext}\ ,~~~
  {\bf A}^3 = \sqrt3 {\bf A}^8 = -\frac{2K_T ge|\kappa_A|^2}
              {q^2+12K_T g_3^2 |\kappa_A|^2} {\bf A}_{\rm ext}\ ,
\end{equation}
and thus
\begin{equation}
  \mbox{\boldmath ${\cal A}$} = \frac{g}{\sqrt{6}g_3}
  {\bf A}_{\rm ext}\ ,~~~
  \mbox{\boldmath ${\cal A}$}^3 =\frac{\sqrt2 e}{3 g_3}
        \frac{q^2}{q^2+12K_T g_3^2 |\kappa_A|^2}
        {\bf A}_{\rm ext}\ ,~~~
  \mbox{\boldmath ${\cal A}$}^8=0\ .
  \label{resmagcfl}
\end{equation}
The term $\mbox{\boldmath ${\cal A}$}^3$, which is proportional to ${\bf
A}_{\rm ext}$, is screened by the Meissner effect within a penetration depth
\begin{equation}
   \lambda_{\rm CFL}\equiv
           \frac{1}{2\sqrt{3 K_T}g_3
                       |\kappa_A|}\ .
   \label{pdcfl}
\end{equation}
[However, see note \cite{fluxdiff}.] The mixed field, $\mbox{\boldmath
${\cal A}$}$, defined by Eq.\ (\ref{mixcfl}), also proportional to ${\bf
A}_{\rm ext}$, propagates freely in the superconductor.  The behavior of the
field $\mbox{\boldmath ${\cal A}$}^3$ is comparable to that in a Type I
superconductor, independent of the relative magnitudes of the coherence length
and the penetration depth.  Due to the absence of phase-gradient induced
current terms, an external field does not lead to vortex lines; a vortex
placed inside the condensate would be topologically unstable.  We note that
since the structure factor $f_{\alpha38}$ vanishes, the gluon self-coupling
term included in the field strength tensor (\ref{glm}) does not appear,
allowing us to generalize expression (\ref{resmagcfl}) straightforwardly to
the case of finite external fields.

    Under the external field ${\bf H}_{\rm ext} = \nabla\times {\bf A}_{\rm
ext}$, the free energy difference near $T_c$, Eq.\ (\ref{domega2}), reduces to
\begin{equation}
   \Delta\Omega = 3{\bar\alpha}|\kappa_A|^2
                  + 9{\bar\beta}_{\rm CFL}|\kappa_A|^4
                  + 6K_T (|\nabla \kappa_A|^2
                  + g_3^2 |\mbox{\boldmath ${\cal A}$}^3|^2|\kappa_A|^2)
                  +\frac 12 \left(|\mbox{\boldmath ${\cal B}$}^3|^2
                  + |\mbox{\boldmath ${\cal B}$}|^2
                  - |{\bf H}_{\rm ext}|^2\right)\ ,
   \label{omegacfl}
\end{equation}
where
\begin{equation}
   \mbox{\boldmath ${\cal B}$} \equiv
         \nabla\times\mbox{\boldmath ${\cal A}$}\ ,~~~
   \mbox{\boldmath ${\cal B}$}^3 \equiv
         \nabla\times \mbox{\boldmath ${\cal A}$}^3
\end{equation}
are the Abelian field strengths.  The gap equation (cf. (\ref{gapeqnkappa}))
follows directly from Eq.~(\ref{omegacfl}).

    The system remains a superfluid for magnetic fields $H_{\rm ext}$ below a
critical field $H_c$.  For $H_{\rm ext}<H_c$, the equilibrium system expels
the part of the external field that leads to $\mbox{\boldmath ${\cal B}$}^3$.
An external field greater than $H_c$ entirely penetrates the system, and the
system becomes normal.  The favored phase is the one that minimizes the Gibbs
free energy density, $G= \Omega -{\bf H}_{\rm ext}\cdot{\bf B}$, at constant
external field, ${\bf H}_{\rm ext}$.  Since $\mbox{\boldmath ${\cal B}$}^3$
is screened out in the color-flavor locked phase in bulk,
\begin{equation}
  G_s = F_n - \frac{{\bar\alpha}^2}{4{\bar\beta}_{\rm CFL}}
     +\frac12|\mbox{\boldmath ${\cal B}$}|^2
     - {\bf H}_{\rm ext}\cdot{\bf B}
   = F_n - \frac{{\bar\alpha}^2}{4{\bar\beta}_{\rm CFL}}
         -\frac{g^2}{12g_3^2} |{\bf H}_{\rm ext}|^2\ .
\end{equation}
The normal phase Gibbs free energy density is
\begin{equation}
   G_n = F_n -\frac 12 |{\bf H}_{\rm ext}|^2\ ,
\end{equation}
so that the difference of Gibbs free energy densities of the superfluid and
normal phases is
\begin{equation}
  \Delta G = - \frac{{\bar\alpha}^2}{4{\bar\beta}_{\rm CFL}}
             +\Delta G_{\rm mag}
\end{equation}
with the magnetic part
\begin{eqnarray}
   \Delta G_{\rm mag} &=&
             \frac 12 |\mbox{\boldmath ${\cal B}$}|^2
             -{\bf H}_{\rm ext}\cdot{\bf B}
             +\frac 12 |{\bf H}_{\rm ext}|^2
   \nonumber \\
    &=&  \frac{e^2}{9g_3^2} |{\bf H}_{\rm ext}|^2\ .
  \label{magcfl}
\end{eqnarray}
This free energy difference vanishes at the critical field
\begin{equation}
   H_{\rm c} = \frac{3 g_3|\bar\alpha|}
                 {2e\sqrt{{\bar\beta}_{\rm CFL}}}
             =\frac{3}{2e\xi_{\rm CFL}\lambda_{\rm CFL}}\ .
  \label{hccfl}
\end{equation}

    The coherence length (\ref{clcfl}), the penetration depth (\ref{pdcfl}),
and the critical field (\ref{hccfl}) can be estimated from Eqs.\
(\ref{alpha})--(\ref{beta}) and (\ref{K}) in the limit of weak coupling as
\begin{equation}
   \xi_{\rm CFL} \simeq 0.26 \left(\frac{100 {\rm ~MeV}}{T_c}\right)
              \left(1-\frac{T}{T_c}\right)^{-1/2}
                             {\rm ~fm}\ ,
  \label{clisrough}
\end{equation}
\begin{equation}
   \lambda_{\rm CFL} \simeq 2.1 \left(\frac{1}{g_3}\right)
               \left(\frac{300 {\rm ~MeV}}{\mu/3}\right)
              \left(1-\frac{T}{T_c}\right)^{-1/2}
              {\rm ~fm}\ ,
  \label{lengthsroughcfl}
\end{equation}
and
\begin{equation}
  H_c \simeq 1.8
  \times10^{19} g_3  \left(\frac{T_c}{100 {\rm ~MeV}}\right)
               \left(\frac{\mu/3}{300 {\rm ~MeV}}\right)
               \left(1-\frac{T}{T_c}\right)
                {\rm ~G}\ .
  \label{hccflrough}
\end{equation}
In these expressions, $T_c$ and $\mu$ are normalized by their typical
values at lower densities.  Although strictly valid near $T_c$, these
estimates of the characteristic lengths and critical field can be extrapolated
to $T\ll T_c$ with the given temperature dependence in terms of $1-T/T_c$
without modification, an approximation which should be accurate to within a
factor of two, as in the case of ordinary superconductors \cite{dG1}.  The
estimate (\ref{hccflrough}) indicates that a color-flavor locked condensate,
if present in a neutron star core, would not be destroyed by ambient neutron
star magnetic fields.

\subsection{Response to rotation}
\label{subsec:CFLROT}

    We turn to analyzing the response of a color-flavor locked homogeneous
condensate near $T_c$ to rotation.  For simplicity, we consider the condensate
to be in thermodynamic equilibrium in a cylindrical vessel rotating with
constant angular velocity $\mbox{\boldmath $\omega$}$.  In this situation, the
normal component corotates with the vessel as a rigid body.  The superfluid
component, however, behaves independently; to examine its properties, we
transform to the rotating frame in which the container walls are at rest,
assuming, as in a neutron star, that the velocities associated with rotation
to be much smaller than the speed of light and describing the flow properties
nonrelativistically.

    The free energy density in the rotating frame is given by $\Omega -
\mbox{\boldmath $\omega$}\cdot{\bf L}$, where ${\bf L}$ is the angular
momentum density, given in terms of the order parameter by
\begin{equation}
    {\bf L} = {\bf r}\times {\bf g}_s,
 \label{angmom}
\end{equation}
with ${\bf g}_s$ given by Eq.~(\ref{gs}).  We neglect here the normal
component contribution to the angular momentum.  Using Eq.~(\ref{phicfl1}),
we find that the superfluid mass density, (\ref{josephson}), is
\begin{equation}
  \rho_s = \frac{16}{3} K_T \mu^2|\kappa_A|^2,
  \label{rhoscfl}
\end{equation}
and
\begin{equation}
  {\bf g}_s  =  4iK_T \mu (\kappa_A^*\nabla\kappa_A-
                       \kappa_A\nabla\kappa_A^*) \ .
\end{equation}
Thus in the rotating frame,
the gradient energy included in Eq.\ (\ref{omegacfl}) becomes
\begin{equation}
 \Omega_{g} = 6K_T\{ |\nabla \kappa_A|^2
                -(2i\mu /3)\mbox{\boldmath $\omega$}\cdot
                [{\bf r}\times(\kappa_A^*\nabla\kappa_A-
                       \kappa_A\nabla\kappa_A^*)]
             +g_3^2 |\mbox{\boldmath ${\cal A}$}^3|^2 |\kappa_A|^2\}\ ,
   \label{omegagcfl}
\end{equation}
with $\mbox{\boldmath ${\cal A}$}^3$ given by Eq.\ (\ref{mix3cfl}).  Note
that the gradient term is independent of the vector potentials; thus rotation
produces a lattice of vortex lines, with the coarse grained superflow pattern
simulating corotation of the condensate.  In contrast, in an isoscalar
condensate, as we discuss in Sec.\ \ref{subsec:ISROT}, a rotating condensate
produces a London magnetic field rather than rotational vortices.  We assume
in the following zero external magnetic field.

    A singly quantized rotational vortex centered on the cylinder axis has the
structure,
\begin{equation}
  \kappa_A = e^{-i\phi}|\kappa_A|\ ,
  \label{rotvort}
\end{equation}
where $\phi$ is the azimuthal angle around the vortex line.  The
corresponding baryon current density, defined by Eq.\ (\ref{gs}), is
\begin{equation}
  {\bf j}_s= 8K_T|\kappa_A|^2 \nabla \phi
  = \frac32  \frac{n_s}{\mu} \nabla \phi = n_s {\bf v}_s \ ,
   \label{jscfl}
\end{equation}
where we relate $\kappa_A$ to $n_s$ via Eq.~(\ref{rhoscfl}).  Integrating
${\bf v}_s$ around a closed loop surrounding the vortex line, we have
\begin{equation}
   \oint d{\bf \ell} \cdot {\bf v}_s= 2\pi \frac{3}{2\mu}\ ;
\end{equation}
thus an individual singly quantized vortex has circulation $3\pi/\mu$.
Specializing to a circular contour of radius $r$ centered on the line, we see
that the vortex velocity, in the azimuthal direction, is given by
\begin{equation}
  |{\bf v}_s| = \frac{3}{2\mu r}\ ,
    \quad \mbox{for $r > \xi_{\rm CFL}$}\ .
     \label{currentcfl}
\end{equation}

    In addition the energy per unit length of an isolated vortex is
\begin{equation}
 T_L = \frac{9\pi n_s}{4\mu}
     \ln\left(\frac{R}{\xi_{\rm CFL}}\right)\ ,
    \label{linecfl}
\end{equation}
where $R$ is the container radius.  The onset of vorticity occurs at the
critical rotation rate $\omega_{c1}$, where the energy in the rotating frame
$T_L/\pi R^2 - \mbox{\boldmath $\omega$}\cdot {\bf L}$ first vanishes; thus
\begin{equation}
   \omega_{c1} = \frac{3}{2\mu R^2}
            \ln\left(\frac{R}{\xi_{\rm CFL}}\right)\ .
   \label{omegac}
\end{equation}

    At finite rotation speeds $\gg \omega_{c1}$ the system forms a triangular
lattice of singly quantized vortices.  The overall flow of the condensate is
essentially that of a solid body at angular velocity $\omega$, with
velocity, Eq.~(\ref{currentcfl}), close to the vortex cores.  This situation
is similar to the case of superfluid neutron vortices in a rotating neutron
star where the quantization of circulation is $\pi/\mu$ \cite{bpp}.  The net
circulation around the cylinder is $3\pi N_v/\mu$, where $N_v$ is the total
number of the vortices.  Since at the boundary, $|{\bf v}_s| \simeq \omega R$,
the circulation is $2\pi R^2 \omega$.  Thus
\begin{eqnarray}
   N_v &=& \frac23 \mu R^2 \omega
   \nonumber \\
       &\simeq& 6.4\times 10^{18}
        \left(\frac{1 {\rm ~ms}}{P_{\rm rot}}\right)
        \left(\frac{\mu/3}{300 {\rm ~MeV}}\right)
        \left(\frac{R}{10 {\rm ~km}}\right)^2\ ,
    \label{nv}
\end{eqnarray}
where in the latter equation the quantities $P_{\rm
rot}(\equiv2\pi/\omega)$, $\mu/3$, and $R$ are normalized to characteristic
neutron star scales.  For these values, the intervortex spacing, $\sim(\pi
R^2/N_v)^{1/2}$, is much larger than the Ginzburg-Landau coherence length
$\xi_{\rm CFL}$, except immediately near $T_c$.

    The total angular momentum, ${\bf L}_{\rm tot}$, of a vortex line, the
integral of $\mu {\bf r}\times{\bf j}_s$, is
\begin{equation}
   {\bf L}_{\rm tot}=\frac32 N_s\hat z\ ,
   \label{L}
\end{equation}
where $\hat z$ is the direction of the vortex axis, and $N_s$ is the total
superfluid baryon number.  As in a neutron superfluid in a neutron star, a
lattice of vortices in a color-flavor locked superfluid is an effective store
of angular momentum.

    To calculate the superflow energy density, $\Omega_{\rm rot}$, of a vortex
lattice in the rotating frame we note that in the logarithmic integral for the
kinetic energy per line, the radial integration is cutoff at the vortex lattice
constant $a$, so that the line energy is given by Eq.~(\ref{linecfl}) only
with $R$ in the logarithm replaced by $a =(2\sqrt3\pi/\mu\omega)^{1/2}$; thus
in the rotating frame,
\begin{equation}
   \Omega_{\rm rot} = \Delta \Omega_{\rm rot}
      - \frac12 \rho_s ({\mbox{\boldmath $\omega$}}\times {\bf r})^2\ ,
\end{equation}
where the latter term is the free energy in the rotating frame of the
(coarse grained) corotation of the fluid, and
\begin{equation}
 \Delta \Omega_{\rm rot} \simeq \frac{3 n_s}{2} \omega
           \ln\left(\frac{a}{\xi_{\rm CFL}}\right)
   \label{domegarot}
\end{equation}
is the local vortex contribution.  In weak coupling, $\Delta\Omega_{\rm
rot}$ can be estimated from Eqs.\ (\ref{alpha})--(\ref{beta}) and (\ref{K}) as
\begin{equation}
   \Delta\Omega_{\rm rot} \sim 3\times10^{-18}
    \left(\frac{\mu/3}{300 {\rm ~MeV}}\right)^3
    \left(\frac{1 {\rm ~ms}}{P_{\rm rot}}\right)
    \left(1-\frac{T}{T_c}\right)
    \ln\left(\frac{a}{\xi_{\rm CFL}}\right)
    {\rm ~MeV} {\rm ~fm^{-3}}\ .
   \label{domegarotrough}
\end{equation}
The upper critical velocity, $\omega_{c2}$, at which rotation destroys
superfluidity, occurs where the intervortex spacing approaches the coherence
length, an impossibly high velocity to be relevant for neutron stars.

\section{Isoscalar, color-antitriplet channel}
\label{sec:IS}

    The response of the isoscalar, color-antitriplet phase to magnetic fields
and rotation is quite different from that of the color-flavor locked phase.
Magnetic fields lead to topologically stable vortices, as in laboratory Type
II superconductors, while rotation of the condensate induces a London magnetic
field \cite{london,varenna}, rather than rotational vortices.  Since these
external disturbances do not distinguish any direction in color space, the
color structure of the order parameters is unaffected.  We specialize, without
loss of generality, to a gap matrix having the specific color orientation:
\begin{equation}
  ({\bf d}_a)_i \equiv e^{i\varphi_0}|{\bf d}| \delta_{aB}\delta_{is}\ ,
  \label{bs}
\end{equation}
where $B$ denotes the blue and $s$ the strange directions; then
\begin{equation}
   (\phi_{+})_{abij} = \epsilon_{ijs}\epsilon_{abB} e^{i\varphi_0}|{\bf d}|.
  \label{bsphi}
\end{equation}
We concentrate on the London limit in which the spatial gradient of the
magnitude of the gap $|{\bf d}|$ can be neglected (see Eq.~(\ref{deltakappa}))
and that of the phase $\varphi_0$ varies very slowly.  This approximation is
sufficient to describe the supercurrent structure outside the vortex cores.

\subsection{Response to magnetic fields}
\label{subsec:ISMAG}

    Substitution of the gap (\ref{bsphi}) into Eqs.\ (\ref{cmax}) and
(\ref{emmax}) leads to the relations between the supercurrent densities and
the vector potentials,
\begin{equation}
   {\bf J}^\alpha=0\ ,~~~ \alpha=1,2,3\ ,
   \label{j1is}
\end{equation}
\begin{equation}
   {\bf J}^4 = -K_T g |{\bf d}|^2 (g{\bf A}^4 - 2{\rm Im}\nabla {\hat d}_R)\ ,
   \label{j4is}
\end{equation}
\begin{equation}
   {\bf J}^\alpha = -K_T g^2 |{\bf d}|^2 {\bf A}^\alpha\ ,~~~ \alpha=5,7\ ,
   \label{j5is}
\end{equation}
\begin{equation}
   {\bf J}^6 = -K_T g |{\bf d}|^2 (g{\bf A}^6 - 2{\rm Im}\nabla {\hat d}_G)\ ,
   \label{j6is}
\end{equation}
\begin{equation}
   {\bf J}^8 = -\frac{4}{\sqrt3} K_T g |{\bf d}|^2 \left(\nabla \varphi_0
           +\frac{g}{\sqrt3}{\bf A}^8 + \frac{e}{3} {\bf A}\right)\ ,
   \label{j8is}
\end{equation}
\begin{equation}
   {\bf J} = \frac{e}{\sqrt3 g}{\bf J}^8 \ ,
   \label{jis}
\end{equation}
where ${\bf {\hat d}}\equiv{\bf d}/|{\bf d}|$.

    As in the color-flavor locked case, it is useful to diagonalize these
relations via the linear combinations \cite{gorbar}
\begin{equation}
 \mbox{\boldmath ${\cal J}$}
   \equiv \frac{\sqrt3 g {\bf J} - e {\bf J}^8}{3g_8}\ ,~~~
 \mbox{\boldmath ${\cal A}$}
   \equiv \frac{\sqrt3 g {\bf A} - e {\bf A}^8}{3g_8}\ ,
   \label{mixis}
\end{equation}
\begin{equation}
 \mbox{\boldmath ${\cal J}$}^8
   \equiv \frac{\sqrt3 g {\bf J}^8 + e {\bf J}}{3g_8}\ ,~~~
 \mbox{\boldmath ${\cal A}$}^8
   \equiv \frac{\sqrt3 g {\bf A}^8 + e {\bf A}}{3g_8}\ ,
   \label{mix8is}
\end{equation}
where
\begin{equation}
 \mbox{\boldmath ${\cal J}$} \equiv 0
   \label{mixjis}
\end{equation}
and
\begin{equation}
   \mbox{\boldmath ${\cal J}$}^8 = -4 K_T g_8 |{\bf d}|^2 \left(
   \nabla \varphi_0
             + g_8 \mbox{\boldmath ${\cal A}$}^8\right)\ ,
   \label{mixj8is}
\end{equation}
with
\begin{equation}
   g_8 = \frac 13 \sqrt{3g^2+e^2}
\end{equation}
the coupling constant associated with the field $\mbox{\boldmath ${\cal
A}$}^8$.  Thus five of the nine gauge fields are Meissner screened, as
expected from the results for color Meissner effects on virtual gluons
mediating quark-quark interactions (see, e.g., Ref.\ \cite{dirk1}).
Expressions (\ref{mixjis}) and (\ref{mixj8is}) indicate that the photonic
gauge field, $\mbox{\boldmath ${\cal A}$}$, modified by a gluonic component of
charge $\alpha=8$, is a free field, whereas the field $\mbox{\boldmath ${\cal
A}$}^8$ couples to the gradient of the phase $\varphi_0$.  This latter
coupling underlies the response of the isoscalar homogeneous condensate to
external magnetic fields and rotation.

    Equations (\ref{res1}) and (\ref{res2}) determine the linear response of
the system to weak external magnetic fields.  For the order parameter
(\ref{bsphi}), an external electromagnetic vector potential leads to terms in
these equations proportional to $\lambda^{\alpha}_{BB}=\lambda^{\alpha}_{33}$,
which in the standard representation is non-zero only for $\alpha = 8$.  Thus
only a field ${\bf A}^8$ is produced; the fields ${\bf A}^\alpha$ remain zero
for $\alpha=1$--7.  The absence of the $\alpha=4$ and 6 gauge fields implies
via Eqs.~(\ref{j4is}) and (\ref{j6is}) that in equilibrium external magnetic
fields do not lead to non-zero gradients ${\rm Im}\nabla {\hat d}_R$ and
${\rm Im}\nabla {\hat d}_G$.  For the electromagnetic and $\alpha=8$
gauge fields we obtain
\begin{equation}
  {\bf A}=\frac{q^2+(4/3)K_T g^2|{\bf d}|^2}
          {q^2+4K_T g_8^2|{\bf d}|^2}
          {\bf A}_{\rm ext}\ ,~~~
  {\bf A}^8 = - \frac{(4/3\sqrt3)K_T ge|{\bf d}|^2}
                   {q^2 + 4 K_T g_8^2|{\bf d}|^2} {\bf A}_{\rm ext}\ ,
\end{equation}
or, equivalently,
\begin{equation}
  \mbox{\boldmath ${\cal A}$} = \frac{g}{\sqrt3 g_8}
  {\bf A}_{\rm ext}\ ,~~~
  \mbox{\boldmath ${\cal A}$}^8 = \frac{e}{3g_8}
  \frac{q^2}{q^2+4K_T g_8^2|{\bf d}|^2} {\bf A}_{\rm ext}\ .
  \label{resmagis}
\end{equation}
We note two features similar to the case of the color-flavor locked
condensate.  First, the modified field $\mbox{\boldmath ${\cal A}$}^8$ is
subject to the Meissner effect with a penetration depth \cite{BS}
\begin{equation}
     \lambda_{\rm IS}=
           \frac{1}{2 g_8 \sqrt{K_T} |{\bf d}|}\ ;
  \label{pdis}
\end{equation}
the modified field $\mbox{\boldmath ${\cal A}$}$ is free.  Second, the
gluon self-coupling term in the field strength tensor (\ref{glm}) is absent.
This fact allows the extension of Eq.\ (\ref{resmagis}) to finite external
fields.

     The free energy difference near $T_c$, Eq.\ (\ref{domega2}), may
now be written in terms of $\mbox{\boldmath ${\cal A}$}$ and
$\mbox{\boldmath ${\cal A}$}^8$ as
\begin{equation}
   \Delta\Omega ={\bar\alpha}|{\bf d}|^2 + {\bar\beta}_{\rm IS}|{\bf d}|^4
                  + 2K_T |(\partial_l + ig_8 {\cal A}^8_l){\bf d}|^2
                  +\frac 12 |\mbox{\boldmath ${\cal B}$}^8|^2
                  +\frac 12 |\mbox{\boldmath ${\cal B}$}|^2
                  -\frac 12 |{\bf H}_{\rm ext}|^2\ ,
   \label{omegais}
\end{equation}
where
\begin{equation}
   \mbox{\boldmath ${\cal B}$} \equiv
         \nabla\times\mbox{\boldmath ${\cal A}$}\ ,~~~
   \mbox{\boldmath ${\cal B}$}^8 \equiv
         \nabla\times \mbox{\boldmath ${\cal A}$}^8
\end{equation}
are the Abelian field strengths, and $\bar\beta_{\rm IS}=\beta_1 + \beta_2$.
As in ordinary superconductors \cite{dG1}, we identify the
Ginzburg-Landau coherence length, $\xi_{\rm IS}$, from expression
(\ref{omegais}) as
\begin{equation}
   \xi_{\rm IS} = \left(2K_T/|{\bar\alpha}|\right)^{1/2}\ ,
  \label{clis}
\end{equation}
a result identical with the color-flavor locked expression (\ref{clcfl}).

    The thermodynamic critical field, $H_c$, associated with ${\bf B}=\nabla
\times{\bf A}$, is the field at which the Gibbs free energy of the normal
state drops to that of the superconducting state.  Since $\mbox{\boldmath
${\cal B}$}^8$ is screened out in the isoscalar phase in bulk,
\begin{equation}
  G_s = F_n - \frac{{\bar\alpha}^2}{4{\bar\beta}_{\rm IS}}
     +\frac12|\mbox{\boldmath ${\cal B}$}|^2
     - {\bf H}_{\rm ext}\cdot{\bf B}
   =F_n - \frac{{\bar\alpha}^2}{4{\bar\beta}_{\rm IS}}
         -\frac{g^2}{6g_8^2} |{\bf H}_{\rm ext}|^2\ ,
\end{equation}
where we have used the identity
\begin{equation}
  {\bf B} = \frac{e}{3g_8}\mbox{\boldmath ${\cal B}$}^8 +
  \frac{g^2}{3g_8^2}{\bf H}_{\rm ext},
\end{equation}
derived from Eqs.~(\ref{mixis}), (\ref{mix8is}) and the first of
(\ref{resmagis}).  The difference in Gibbs free energy densities of the paired
and normal phase is thus
\begin{equation}
  \Delta G = - \frac{{\bar\alpha}^2}{4{\bar\beta}_{\rm IS}}
             +\Delta G_{\rm mag},
\end{equation}
with the magnetic part
\begin{eqnarray}
   \Delta G_{\rm mag} &=&
             \frac 12 |\mbox{\boldmath ${\cal B}$}|^2
             -{\bf H}_{\rm ext}\cdot{\bf B}
             +\frac 12 |{\bf H}_{\rm ext}|^2
  \nonumber \\
     &=&  \frac{e^2}{18g_8^2} |{\bf H}_{\rm ext}|^2\ .
   \label{magis}
\end{eqnarray}
Then,
\begin{equation}
   H_{\rm c} = \frac{3g_8}{e} \frac{|\bar\alpha|}
                 {\sqrt{2{\bar\beta}_{\rm IS}}}
              = \frac{3}{\sqrt2 e\xi_{\rm IS}\lambda_{\rm IS}}\ .
  \label{hcis}
\end{equation}

    The presence of the gauge field, $\mbox{\boldmath ${\cal A}$}^8$, in the
derivative term in the free energy indicates that the system can support
magnetic vortices, as in Type II laboratory superconductors.  Whether or not
vortices appear depends, as we discuss below, on the relative size of the
screening and coherence lengths.  Let us first analyze the structure outside
the core of a single vortex, whose core, of radius $\sim\xi_{\rm IS}$, we take
to be aligned along the z-axis.  Then the order parameter of the vortex is
given by Eq.~(\ref{bs}), with $\varphi_0 = -\phi$, where $\phi$ is, as before,
the azimuthal angle around the line.  The field equation for $\mbox{\boldmath
${\cal B}$}^8$, derived from extremization of the integral of Eq.\
(\ref{omegais}) over the system with respect to $\mbox{\boldmath ${\cal
A}$}^8$ at fixed $\mbox{\boldmath ${\cal A}$}$, is
\begin{equation}
   \mbox{\boldmath ${\cal B}$}^8 + \lambda_{\rm IS}^2 \nabla\times
    (\nabla\times \mbox{\boldmath ${\cal B}$}^8) =
     \frac{1}{g_8}\nabla\times\nabla\phi\
    = \frac{2\pi}{g_8}\delta(x)\delta(y) {\hat z}\ ,
  \label{max8}
\end{equation}
in the London limit in which the spatial gradient of the magnitude of the
gap $|{\bf d}|$ is negligible.  The condition of quantization of
magnetic flux, found from integrating Eq.~(\ref{max8}) over the interior of
a contour in the x-y plane surrounding the line, is
\begin{equation}
   \oint d{\bf \ell}\cdot (\mbox{\boldmath ${\cal A}$}^8
                 +\lambda_{\rm IS}^2\mbox{\boldmath ${\cal J}$}^8)
      = \frac{2\pi}{g_8}\ .
  \label{quanti}
\end{equation}
The flux quantum, $\phi_8$, is thus $2\pi/g_8$; note that in the absence
of color coupling ($g=0$) the flux quantum reduces to $6\pi/e$, which is
$2\pi$ divided by the net charge, $2/3 - 1/3$, of the pair.

    As in ordinary Type II superconductors \cite{dG1}, it is straightforward
to solve the London equation (\ref{max8}).  The current density
$\mbox{\boldmath ${\cal J}$}^8$, which flows in the azimuthal direction around
the line, has a magnitude
\begin{equation}
  |\mbox{\boldmath ${\cal J}$}^8| =
 \left\{
 \begin{array}{ll}
  \displaystyle{\frac{\phi_8}{2\pi\lambda_{\rm IS}^2 r}}\ ,
   & \quad \mbox{for $\xi_{\rm IS} < r \ll \lambda_{\rm IS}$}\ , \\
  \displaystyle{\frac{\phi_8}{2\pi\lambda_{\rm IS}^2}
                \left(\frac{\pi\lambda_{\rm IS}}{2r}\right)^{1/2}
       \left(\frac{1}{\lambda_{\rm IS}}
             +\frac{1}{2r}\right)}e^{-r/\lambda_{\rm IS}}\ ,
   & \quad \mbox{for $r \gg \lambda_{\rm IS}$}\ , \\
 \end{array}
 \right.
     \label{currentis}
\end{equation}
and the current $\mbox{\boldmath ${\cal J}$}^8$ is screened by the
Meissner effect far from the vortex core ($r \gg \lambda_{\rm IS}$); there the
gauge field $\mbox{\boldmath ${\cal A}$}^8$ alone fulfills the quantization
condition (\ref{quanti}).  The vortex line energy per unit length, the sum of
the magnetic and flow energies, is
\begin{equation}
 T_L = \frac{\phi_8^2}{4\pi\lambda_{\rm IS}^2}
     \ln\left(\frac{\lambda_{\rm IS}}{\xi_{\rm IS}}\right)\ .
  \label{tlis}
\end{equation}

    We proceed to classify the superconducting properties according to whether
or not the isoscalar condensate allows magnetic vortices associated with the
field $\mbox{\boldmath ${\cal A}$}^8$ to form, i.e., whether it is Type II or
Type I.  Blaschke and Sedrakian \cite{BS} addressed this problem in the weak
coupling Ginzburg-Landau formalism.  We give the argument here for arbitrary
coupling.  The basic idea is to calculate the energy per unit area,
$\sigma_s$, needed to form a planar surface separating the normal and
superconducting material.  At the thermodynamic critical field $H_c$, applied
parallel to the surface, the surface is in mechanical equilibrium.  For a
surface perpendicular to the z-axis, the surface energy $\sigma_s$ per unit
area may be written as the integral over $z$ of the difference of the total
Gibbs free energy density, $\Delta\Omega(z)|_{H_{\rm ext}\to H_c}+F_n
+\frac12 H_c^2-H_c|{\bf B}(z)|$, and the value $F_n-\frac12
H_c^2$ for the limiting case in which the thickness of the interface vanishes
($\lambda_{\rm IS} \to 0$ and $\xi_{\rm IS} \to 0$); we find
\begin{eqnarray}
   \sigma_s &=& \int_{-\infty}^{\infty}dz\left[\Delta\Omega(z)|_{H_{\rm ext}
                        \to H_c}- H_c(|{\bf B}(z)|-H_c)\right]
   \nonumber \\
            &=& \int_{-\infty}^{\infty}dz \left\{ \frac{}{}
                 {\bar\alpha}|{\bf d} (z)|^2
                 + {\bar\beta}_{\rm IS}|{\bf d} (z)|^4
   \right. \nonumber \\  & &
                + 2K_T |[\partial_l + ig_8 {\cal A}^8_l(z)]{\bf d}(z)|^2
   \nonumber \\ & & \left.
                +\frac 12 \left[
                  |\mbox{\boldmath ${\cal B}$}^8(z)|
                  -\frac{e}{3g_8} H_c\right]^2 \right\} \nonumber \\
                & \sim &
          \frac{\xi_{\rm IS}}{2}\left(\frac{e}{3g_8}H_c\right)^2
                (1-\sqrt2\kappa_{\rm IS})\ ,~~~
                \mbox{for $\kappa_{\rm IS} \sim 1/\sqrt2$}\ ,
  \label{sigmasis}
\end{eqnarray}
where the Ginzburg-Landau parameter $\kappa_{\rm IS}$ is defined by
\begin{equation}
   \kappa_{\rm IS} \equiv \lambda_{\rm IS}/\xi_{\rm IS}\ ,
  \label{kappa}
\end{equation}
with $\lambda_{\rm IS}$ and $\xi_{\rm IS}$ given by Eqs.\ (\ref{pdis}) and
(\ref{clis}).  The calculation of the prefactor of $\sigma_s$ must be done
numerically.  The structure of Eq.~(\ref{sigmasis}) is identical to that of
ordinary superconductors \cite{dG1} with the surface energy changing sign from
positive to negative as $\kappa_{\rm IS}$ goes through $1/\sqrt2$; thus we are
led to the same criterion, namely that the system is Type I, with no vortex
formation, for $\sigma_s > 0$, and Type II, with an Abrikosov-Schubnikov
vortex phase, for $\sigma_s < 0$.

    While in a Type I superconductor, superconductivity is destroyed by an
external uniform magnetic field $H_{\rm ext}$ greater than $H_c$, in a Type II
superconductor, vortices form at a lower critical field $H_{c1}<H_c$, and
increase in density up to an upper critical field $H_{c2}\,(>H_c)$, where the
system turns normal.  At the field $H_{c2}$, magnetic vortex cores essentially
fuse.  At the lower critical field $H_{c1}$, the energy gain from penetration
of the magnetic field just compensates for the energy loss due to the line
energy $T_L$, Eq.\ (\ref{tlis}).  The energy gain per unit volume from field
penetration, given in terms of the magnetic induction ${\cal B}^8$ and the
corresponding magnetic field ${\cal H}^8 = (e/3g_8)H_{\rm ext}$ (cf. the
second part of Eq.~(\ref{mix8is})), is $E_{\rm mag}=-{\cal B}^8{\cal H}^8$.
Integrating over the volume including the vortex, we find, using
Eq.~(\ref{quanti}), a total energy gain from field penetration, $(e/3g_8)
H_{\rm ext}\phi_8$.  Equating this term to $T_L$ we find that
\begin{equation}
 H_{c1} = \frac{3}{2e\lambda_{\rm IS}^2}
          \ln\left(\frac{\lambda_{\rm IS}}{\xi_{\rm IS}}\right)\ .
   \label{hc1is}
\end{equation}
Just below $H_{c2}$ the pairing gap has a tiny magnitude but strong
spatial variation, and is determined by a linearized and inhomogeneous version
of the gap equation (\ref{gapeq}), as first discussed by Abrikosov
\cite{abrikosov}.  From the condition for the existence of nontrivial
solutions to this equation, we obtain the familiar form
\begin{equation}
   H_{\rm c2} = \frac{3}{e{\xi_{\rm IS}}^2}\ .
  \label{hc2is}
\end{equation}

    At high densities, where the system is weakly coupled, $\kappa_{\rm IS}$
reduces to
\begin{equation}
 \kappa_{\rm IS} = 18\pi^2
                \left(\frac{2}{7\zeta(3)}\right)^{1/2}
                \frac{T_c}{g_8\mu}\ .
   \label{kappawc}
\end{equation}
Since $T_c\sim \mu g^{-5} e^{-3\pi^2/\sqrt2 g}$ \cite{BLR}, $\kappa_{\rm
IS}$ is far smaller than unity, so that the system is Type I near $T_c$.  It
is not out of the question that at low densities where $T_c$ can be $\sim
0.1\, \mu$ and $g_8 \sim 1$, that $\kappa_{\rm IS}$ can become larger than
$1/\sqrt2$ and the system can become Type II \cite{BS}, expelling the field
$\mbox{\boldmath ${\cal B}$}^8$ completely for $H_{\rm ext}<H_{c1}$, while
allowing the external magnetic field $H_{\rm ext}$ to penetrate freely in the
form of the photon-gluon mixed field $\mbox{\boldmath ${\cal B}$}$ of strength
$(g/\sqrt3 g_8)H_{\rm ext}$ [Eq.\ (\ref{resmagis})].

    The magnitude of the coherence length $\xi_{\rm IS}$, (\ref{clis}), is the
same as $\xi_{\rm CFL}$ in Eq.\ (\ref{clisrough}), while the penetration depth
(\ref{pdis}) can be estimated from
\begin{equation}
   \lambda_{\rm IS} \simeq 1.5 \left(\frac{\sqrt3}{g_8}\right)
               \left(\frac{300 {\rm ~MeV}}{\mu/3}\right)
              \left(1-\frac{T}{T_c}\right)^{-1/2}
              {\rm ~fm}\ .
  \label{pdisrough}
\end{equation}
To estimate the magnitude of the critical fields (\ref{hcis}),
(\ref{hc1is}), and (\ref{hc2is}) near $T_c$ we use (\ref{alpha})--(\ref{beta})
and (\ref{K}) to find
\begin{equation}
  H_c \simeq 3.6\times10^{19}
  \left(\frac{g_8}{\sqrt3}\right)
   \left(\frac{T_c}{100 {\rm ~MeV}}\right)
               \left(\frac{\mu/3}{300 {\rm ~MeV}}\right)
               \left(1-\frac{T}{T_c}\right)
                {\rm ~G}\ ,
  \label{hcisrough}
\end{equation}
\begin{eqnarray}
  H_{c1} &\simeq& 7.9\times10^{18} 
               \left(\frac{g_8^2}{3}\right)
               \left(\frac{\mu/3}{300 {\rm ~MeV}}\right)^2
               \left[1+\ln\left( \frac{\sqrt3}{g_8}
                \frac{T_c}{100 {\rm ~MeV}} \frac{300 {\rm ~MeV}}{\mu/3}
                     \right)\right]
               \left(1-\frac{T}{T_c}\right)
                {\rm ~G}\ ,
  \label{hc1isrough}
\end{eqnarray}
and
\begin{equation}
  H_{c2} \simeq 2.9 \times10^{20} \left(\frac{T_c}{100 {\rm ~MeV}}\right)^2
               \left(1-\frac{T}{T_c}\right)
               {\rm ~G}\ .
  \label{hc2isrough}
\end{equation}
Extrapolation of expressions (\ref{hcisrough})--(\ref{hc2isrough}) to low
densities and temperatures indicates that the critical fields $H_c$, $H_{c1}$,
and $H_{c2}$ are several orders of magnitude larger than canonical neutron
star surface fields $\sim 10^{12}$ G. However, to assess the actual situations
possible in neutron stars, one must take into account the history of the
expulsion of the magnetic field and the possibility of freezing in of the
magnetic field; see note \cite {fluxdiff}.

\subsection{Response to rotation}
\label{subsec:ISROT}

    To determine the equilibrium properties of a rotating isoscalar
condensate, we assume, as in the color-flavor locked discussion, that it is in
a cylinder rotating at angular velocity $\mbox{\boldmath $\omega$}$.  We first
note that Eqs.~(\ref{josephson}) and (\ref{bsphi}) imply that the superfluid 
mass density of an isoscalar condensate is
\begin{equation}
  \rho_s = \frac{16}{9} K_T \mu^2|{\bf d}|^2;
  \label{rhois}
\end{equation}
in addition, the momentum carried by the condensate is
\begin{equation}
  (g_s)_l  =  \frac43 K_T \mu
  \{{\bf d}^* \cdot(i\partial_l - g_8 {\cal A}^8_l) {\bf d}
           + [(-i\partial_l - g_8 {\cal A}^8_l) {\bf d}^*]\cdot{\bf d}\} \ .
\end{equation}
Thus, in the rotating frame, the gradient energy in Eq.\ (\ref{omegais})
is
\begin{equation}
     \Omega_{g} =
     2K_T  |(\partial_l + ig_8 {\cal A}^8_l){\bf d}|^2
         -\frac{4}{3}K_T\mu  (\mbox{\boldmath $\omega$}\times{\bf r})_l
      \{{\bf d}^* \cdot(i\partial_l - g_8 {\cal A}^8_l) {\bf d}
           + [(-i\partial_l - g_8 {\cal A}^8_l) {\bf d}^*]\cdot{\bf d}\}\ ,
    \label{omegagis}
\end{equation}
where ${\cal A}^8_l$ is given by Eq.\ (\ref{mix8is}).

    From expression (\ref{omegagis}) we find that in equilibrium, rotation
induces a London magnetic field associated with the gluon-photon mixed
potential $\mbox{\boldmath ${\cal A}$}^8$, rather than generating vortex lines
as in a color-flavor locked condensate.  This response is due to the presence
of charge $g_8$, and is thus similar to the case of ordinary rotating
superconductors \cite{london,varenna} and proton superconductors in rotating
neutron stars \cite{sauls}.  Rotational vortices are topologically unstable; a
rotational vortex of the form (\ref{currentcfl}) in the system can be unwound
by bringing the velocity to zero simultaneously generating a gauge field
${\cal A}^8$ to preserving the quantization condition (\ref{quanti}).  To
derive the London field we assume that the system has zero external magnetic
field.  In a rotating superconductor the local current $\mbox{\boldmath ${\cal
J}$}^8$ is given, deeper in than a penetration depth from the surface, by
\begin{equation}
  \mbox{\boldmath ${\cal J}$}^8 =
          \frac32 g_8 n_s \mbox{\boldmath $\omega$}\times{\bf r}\ ,
  \label{lonj88}
\end{equation}
since the pairs, of charge $g_8$ and baryon number 2/3, corotate with the
vessel.  In the absence of a vortex, the quantization condition on the
vorticity becomes
\begin{equation}
   \oint d{\bf \ell}\cdot (\mbox{\boldmath ${\cal A}$}^8
                 +\lambda_{\rm IS}^2\mbox{\boldmath ${\cal J}$}^8)
      = 0\
  \label{quanti0}
\end{equation}
[cf.~(\ref{quanti}) for a contour surrounding a quantized vortex].
Taking a circular contour of fixed distance $r$ from the rotation axis, we
find
\begin{equation}
   \oint d{\bf \ell}\cdot {\mbox{\boldmath ${\cal A}$}^8}
   = \int d^2r {\cal B}^8_z
      = -\frac{4\pi\mu}{3g_8}\omega r^2.
  \label{lond1}
\end{equation}
Thus ${\cal B}^8_z$ is constant and
\begin{equation}
 \mbox{\boldmath ${\cal B}$}^8
  = - \frac{4\mu}{3g_8} {\mbox{\boldmath $\omega$}},\quad
 \mbox{\boldmath ${\cal A}$}^8
  = - \frac{2\mu}{3g_8} {\mbox{\boldmath $\omega$}}\times {\bf r}\ .
  \label{londfield}
\end{equation}
The magnetic field $\mbox{\boldmath ${\cal B}$}$ remains zero.

    Since ${\bf d}$ is a constant in space, the gradient energy
(\ref{omegagis}) reduces to
$ - \frac12 \rho_s ({\mbox{\boldmath
                $\omega$}}\times {\bf r})^2$
so that the total free energy density associated with rotation, measured in the
rotating frame, is
\begin{equation}
   \Omega_{\rm rot} = \Delta\Omega_{\rm rot}
        - \frac12 \rho_s ( \mbox{\boldmath $\omega$} \times {\bf r})^2 \ ,
\end{equation}
where
\begin{equation}
  \Delta\Omega_{\rm rot} = \frac12 |\mbox{\boldmath ${\cal B}$}^8|^2
  \label{domegarotis}
\end{equation}
is the energy density of the London field.

    We make several observations about the London field (\ref{londfield}).
First, it reduces to the conventional London field of a nonrelativistic
electron superconductor with the replacements $g_8/2 \to -e$ and $\mu/3 \to
m_e$.  Second, since $\mbox{\boldmath ${\cal B}$}=0$, we have $|{\bf
B}^8|=(\sqrt3 g/e)|{\bf B}|$, which implies that the London field is primarily
composed of the color field of charge $\alpha=8$.  The order of magnitude of
the London field is
\begin{eqnarray}
   |\mbox{\boldmath ${\cal B}$}^8|
  &\sim& 0.15
    \left(\frac{\sqrt3}{g_8}\right)
    \left(\frac{1 {\rm ~ms}}{P_{\rm rot}}\right)
    \left(\frac{\mu/3}{300 {\rm ~MeV}}\right) {\rm ~G}\ ,
   \label{lon}
\end{eqnarray}
so that at low densities relevant to neutron star interiors, the London
field is thoroughly negligible compared with the typical magnitude
($\sim10^{12}$ G) of the neutron star surface magnetic fields.  Third, in the
presence of a uniform external field $H_{\rm ext}<H_{c1}$, we have
$\mbox{\boldmath ${\cal B}$}=(g/\sqrt3 g_8){\bf H}_{\rm ext}$, while the
London field (\ref{lon}) is unaffected.  For $H_{\rm ext}>H_{c1}$, where
vortices form, we find that the total field $\mbox{\boldmath ${\cal B}$}$
remains the same, while the total field $\mbox{\boldmath ${\cal B}$}^8$ is the
sum of that associated with the vortices, plus the London field, as long as
the vortex separation is large compared with the penetration depth,
$\lambda_{\rm IS}$.  Fourth, the London field increases with increasing
density, because of the power-law increase in $\mu/3$ and the logarithmic 
decrease in $g_8$, leading to an increasing supercurrent responsible for the 
London field.  Finally, we remark that in the present case in which only 
${\bf A}$ and ${\bf A}^8$ are the nonzero gauge fields, no self-coupling of 
the London field occurs.

\section{Conclusion}
\label{sec:conclusion}

    In this paper we have investigated the response of homogeneous superfluid
quark matter in thermodynamic equilibrium near $T_c$ to magnetic fields and
rotation.  Rotational vortices are topologically stable in the color-flavor
locked phase, as are magnetic vortices in the isoscalar phase.  Table
\ref{summary} summarizes the responses and the corresponding corrections to
the free energy difference between the superconducting and normal phases.  We
have ignored several aspects, including the modification of the structure of
the phases close to $T_c$ due to the rotational and magnetic energies, effects
of the finite strange quark mass $m_s$, and boundary effects.

    The energy correction $\Delta G_{\rm mag}$ due to an external uniform
magnetic field $H_{\rm ext}$ increases the energy of a color-flavor locked
condensate more than that of an isoscalar condensate, as one can see by
comparing the color-flavor locked result $(e^2/9g_3^2)H_{\rm ext}^2$ from Eq.\
(\ref{magcfl}) and the isoscalar result $(e^2/18g_8^2) H_{\rm ext}^2$ from
Eq.\ (\ref{magis}), and noting that $g_3^2<2g_8^2$.  In addition, the energy
correction $\Delta\Omega_{\rm rot}$ due to rotation at constant angular
velocity $\omega$, as given by Eq.\ (\ref{domegarot}) in the color-flavor
locked phase and Eq.\ (\ref{domegarotis}) in the isoscalar phase, raises the
energy of the color-flavor locked state more than that of the isoscalar state
at temperatures $1-T/T_c \gtrsim \omega/g_8^2\mu$, while being more favorable
for the color-flavor locked state in a range $1-T/T_c\lesssim \omega/g_8^2
\mu$.  Although interesting in principle, these terms, for expected magnetic
fields and rotation rates of neutron stars, affect the phase diagram only in a
negligible range around $T_c$ and can be ignored.

    The effects of a finite strange quark mass, $m_s$, are also insignificant
for the response properties, because a non-zero $m_s$ does not change the
structure of the gradient energy, but merely modifies its magnitude in the
color-flavor locked phase through the reduction in the Fermi momentum of the
strange quarks, and modifies $T_c$.

    Generally, the boundaries of the quark superfluid, e.g., an interface with
hadronic matter, contain supercurrents induced by the Meissner effect and
order-parameter gradients.  The resultant energy corrections may possibly
compete with the bulk corrections listed in Table \ref{summary}.  Magnetic
structures have been examined for boundaries of various shapes
\cite{ABR2,SBSV}.  On the other hand, the influence of boundaries on
rotational structures remains an interesting problem.  If a rotating neutron
star contains hadronic matter with a neutron superfluid and quark matter with
a color-flavor locked condensate separated by a spherical interface, then the
connection of the neutron vortices to the quark vortices is a complicated and
interesting issue, since the two types of vortices carry a different baryon
number and hence a different unit of circulation at the interface ($3\pi/\mu$
for quarks and $\pi/\mu$ for neutrons).  For example, a singly quantized
vortex continuing through the interface would produce a difference in the
velocity field around the vortex core, which can lead to a Kelvin-Helmholtz
instability of the interface \cite{lord}.  A connection between three singly
quantized neutron vortices, each of circulation $\pi/\mu$, (or a triply
quantized neutron vortex) and a singly quantized quark vortex would be free
from such velocity differences, since the baryon chemical potential is
continuous across the interface.  Neutron vortices at an interface between
superfluid neutron matter and an isoscalar condensate would terminate, while
color currents responsible for the uniform London field would be arranged in
the vicinity of the interface in such a way as to prevent the color fields
from penetrating into the hadronic region.

\section*{Acknowledgments}

    Author KI would like to acknowledge the hospitality of the Department of
Physics of the University of Illinois at Urbana-Champaign, where this work was
initiated.  Author GB is grateful to Professor T. Hatsuda of Tokyo University
for hospitality during his stay in the Department of Physics of the University
of Tokyo, where this work was completed, and to the Ministry of Education,
Culture, Sports, Science, and Technology of Japan (Monkasho) for supporting
this stay.  This work was also supported in part by a Grant-in-Aid for
Scientific Research provided by Monbusho through Grant No.\ 10-03687, by
National Science Foundation Grant Nos.\ PHY98-00978 and PHY00-98353, and by
RIKEN Special Postdoctoral Researchers Grant No.\ 011-52040.

\appendix
\section*{Homogeneous superfluid}
\label{app:I}

    In this appendix, we summarize the general Ginzburg-Landau analysis for
homogeneous superfluid quark matter of three massless flavors, temperature
$T$, and baryon chemical potential $\mu$, as developed in I, ignoring the
small free energy corrections arising from the constraint of overall color
neutrality.  BCS pairing between two quarks of colors $a$ and $b$, flavors $i$
and $j$, spinor indices $\mu$ and $\nu$, and space-time locations $x$ and $y$,
respectively, is generally characterized by a gap matrix,
$\Delta^{\mu\nu}_{abij}(x-y)$, which is coupled in the action with the quark
spinor $[\psi_{bj}(y)]_{\nu}$ and its charge-conjugate spinor
$[{\bar\psi}^C_{ai}(x)]_{\mu}$, defined by $\psi^{C}_{ai}\equiv C{\bar
\psi}^{T}_{ai}$ with $C=i\gamma^2\gamma^0$ (in the Pauli-Dirac
representation).  The gap $\Delta(x-y)$ is related to the order parameter
$\langle\psi^{C}(x){\bar \psi}(y)\rangle$ via the gap equation [Eq.\ (9) in
I]:
\begin{eqnarray}
  \Delta(k)&=&ig^{2}T\sum_{n~ {\rm odd}}\int\frac{d^{3}q}{(2\pi)^{3}}
 \int d^{4}(x-y)e^{iq(x-y)}\gamma^{\mu}\frac{(\lambda^{\alpha})^{T}}{2}
 \nonumber \\ & & \times
 \left\{\langle
  T[\psi^{C}(x){\bar\psi}(y)]\rangle\Gamma_{\nu\beta}^{(11)}(q,k)
 +\langle T[\psi^{C}(x){\bar \psi}^{C}(y)]\rangle\Gamma_{\nu\beta}^{(21)}(q,k)
  \right\}D_{\mu\nu}^{\alpha\beta}(k-q)\ ,
\end{eqnarray}
where $\Gamma^{(11)}$ is the full quark-quark-gluon vertex,
$\Gamma^{(21)}$ is the full antiquark-quark-gluon vertex, $D$ is the full
gluon propagator, and the Matsubara frequencies are given by $q_0=in\pi T$.
(Note that the convention for this order parameter is the complex conjugate of
the usual condensed matter convention, $\Delta\sim\langle\psi\psi\rangle$.)

    The gap matrix for pairing between quarks with zero total angular
momentum, even parity, and aligned chirality has the structure [I]
\begin{eqnarray}
  \Delta(k) &\equiv& \int d(x-y)\Delta(x-y)e^{ik(x-y)}
  \nonumber \\ &=&
  \gamma^{5}[\phi_{+}(k_{0},|{\bf k}|)\Lambda^{+}({\bf {\hat k}})
  +\phi_{-}(k_{0},|{\bf k}|)\Lambda^{-}({\bf {\hat k}})]\ ,
  \label{delta1}
\end{eqnarray}
where $k$ is the relative four-momentum of the paired quarks, ${\bf {\hat
k}}\equiv {\bf k}/|{\bf k}|$, the
\begin{equation}
  \Lambda^{\pm}({\bf {\hat k}})=
  \frac{1\pm\gamma^{0}\mbox{\boldmath $\gamma\cdot$}{\bf {\hat k}}}{2}
\end{equation}
are energy projection operators for noninteracting massless quarks, and
$\phi_\pm$ denotes the quark-quark or antiquark-antiquark pairing gap.  The
anti-commutation relations for the quark fields require
\begin{equation}
  [\phi_\pm(k_{0},|{\bf k}|)]_{abij}
 =[\phi_\pm(-k_{0},|{\bf k}|)]_{baji}\ .
 \label{pauli}
\end{equation}

    In the weak coupling limit, the coefficients $\alpha^+$, $\beta_1^+$, and
$\beta_2^+$ in the Ginzburg-Landau free energy (\ref{domega}) are
\begin{equation}
  \alpha^+ = N(\mu/3)\ln\left(\frac{T}{T_{c}}\right)\ ,
 \label{alpha}
\end{equation}
\begin{equation}
  \beta_1^+=0\ , \quad
  \beta_2^+=\frac 12 \frac{7\zeta(3)}{8(\pi T_{c})^{2}} N(\mu/3)\ ,
 \label{beta}
\end{equation}
with the ideal gas density of states at the Fermi surface,
\begin{equation}
   N(\mu/3) = \frac{1}{2\pi^{2}}\left(\frac{\mu}{3}\right)^{2}\ .
\end{equation}
Effects beyond weak coupling lead to a nonzero $\beta_1^+$ through the
dependence of the pairing interactions on the gap, and as well modify all the
coefficients (\ref{alpha})--(\ref{beta}) mainly through the normal medium
corrections to $T_c$ and $N(\mu/3)$; however, these coefficients remain to be
determined in the strong coupling regime.

    For condensates antisymmetric in color and flavor, characterized by Eq.\
(\ref{phi1}), substitution of (\ref{phi1}) into $\Delta\Omega^{(0)}$,
Eq.~(\ref{domega}), leads to
\begin{eqnarray}
   \Delta\Omega^{(0)} &=&   {\bar\alpha}\lambda
         +(\beta_{1}+\beta_{2}\Upsilon)\lambda^{2}
   \label{domega3}
\end{eqnarray}
with
\begin{equation}
   {\bar\alpha}\equiv4\alpha^{+}\ , ~~
   \beta_{1}\equiv16\beta_{1}^{+}+2\beta_{2}^{+}\ , ~~
   \beta_{2}\equiv2\beta_{2}^{+}\ ,
\end{equation}
\begin{equation}
   \lambda\equiv\sum_{a}|{\bf d}_{a}|^{2}\ , ~~
   \Upsilon\equiv\frac{1}{\lambda^{2}}
    \sum_{ab}|{\bf d}_{a}^{*}\cdot{\bf d}_{b}|^{2}\ .
  \label{zetalambda}
\end{equation}

    In homogeneous systems, we obtain two types of optimal condensates, as
long as the overall fourth-order term in $\Delta\Omega^{(0)}$ is positive.
The first, the two-flavor color superconducting (2SC) phase, characterized
by $\Upsilon=1$, corresponds to an order parameter involving only one
component of flavor-antitriplet states, i.e., satisfying
\begin{equation}
  {\bf d}_R  \parallel  {\bf d}_G  \parallel  {\bf d}_B\ .
 \label{opis}
\end{equation}
For $\Upsilon=1$ we focus on the isoscalar, color-antitriplet pairing
state with
\begin{equation}
  (\phi_{+})_{abij}=\epsilon_{abc}\epsilon_{ijs}({\bf d}_{c})_{s}\ ;
  \label{icat1}
\end{equation}
this state is degenerate in color space under rotation of the color
axes.

    The other homogeneous condensation is described by a set of order
parameters satisfying $\Upsilon=1/3$ or, equivalently,
\begin{equation}
 {\bf d}_{R}^{*}\cdot{\bf d}_{G}={\bf d}_{G}^{*}\cdot{\bf d}_{B}
     ={\bf d}_{B}^{*}\cdot{\bf d}_{R}=0\ , ~~
 |{\bf d}_{R}|^{2}=|{\bf d}_{G}|^{2}=|{\bf d}_{B}|^{2}\ ,
 \label{opcfl}
\end{equation}
corresponding to the color-flavor locked phase \cite{ARW}; the condensate
in this phase is characterized by its symmetry under simultaneous exchange of
color and flavor.  Color-flavor locked states transform into one another
under global $U(1)$ and flavor (or color) rotation.  The simplest among these
states is described by $({\bf d}_{a})_{i}\propto \delta_{ai}$ and $({\bf
d}_{R})_{u}=({\bf d}_{G})_{d}=({\bf d}_{B})_{s} \equiv \kappa_{A}$; the
corresponding gap matrix is given by
\begin{equation}
  (\phi_{+})_{abij}=\kappa_{A}
  (\delta_{ai}\delta_{bj}-\delta_{aj}\delta_{bi})\ .
 \label{phicfl}
\end{equation}

    For a positive overall fourth order term in the free energy
(\ref{domega3}), the transition from the normal to the superfluid state is
second order.  The resultant state is color-flavor locked for $\beta_2>0$,
while isoscalar for $\beta_2<0$.  We remark in passing that the critical
temperature $T_c$ is the same for these two pairing states, which are induced
by the same instability of the normal phase.  Should the coefficient of the
fourth order term, $\beta_1+\beta_2 \Upsilon$, become negative the phase
transition becomes first order, as discussed in I.


\begin{references}
\bibitem{fluxdiff} In neutron stars, where the normal flux diffusion time
is far greater than the time to cool below the superfluid transition
temperature \cite{bpp}, the proton superfluid does not have time to expel an
external magnetic field, and being a Type II superconductor, forms magnetic
vortices, even though the magnetic fields are below the lower critical field,
$H_{\rm c1}$, for vortex penetration.  The expected flux diffusion times here
are similarly of order $R^2/c^2\tau$, where $R$ is the neutron star core
radius, and $\tau$ is the microscopic scattering time.  Thus the flux
diffusion times, of order millenia or longer, are again too long for magnetic
flux expulsion, and one expects that the system must coexist with the fields
metastably.  In systems with color-flavor locked and Type I isoscalar
condensates, magnetic fields would be frozen in an intermediate state composed
of alternating regions of normal material with flux density $H_c$ and
superconducting material exhibiting incomplete Meissner screening, while in a
Type II isoscalar condensed system, the fields are frozen in a lattice of
vortices.
\end{references}
\end{document}